\documentclass[pra,epsfig,rotate,superscriptaddress,showpacs]{revtex4}
\usepackage{graphicx}
\usepackage{epstopdf}
\usepackage{amsfonts}
\usepackage{amssymb}
\usepackage{amsmath}
\usepackage{subfigure}
\usepackage{prettyref}
\usepackage{float}
\usepackage{color,CJK}
\usepackage{amsmath}
\usepackage{amssymb}
\usepackage{esint}

\usepackage{braket}

\def\be{\begin{equation}}
\def\ee{\end{equation}}
\def\ba{\begin{eqnarray}}
\def\ea{\end{eqnarray}}
\def\iv{{\rm  In}}

\begin{document}
\begin{CJK}{UTF8}{gbsn}
\title
{Lorentz quantum mechanics}

\author{Qi Zhang(张起)}
\affiliation{College of Science, Zhejiang University of Technology,
Hangzhou 310023, China}
\author{Biao Wu(吴飙)} 
      \affiliation{International Center for Quantum Materials, School of Physics, Peking University, Beijing 100871, China}
      \affiliation{Collaborative Innovation Center of Quantum Matter, Beijing 100871, China}
      \affiliation{Wilczek Quantum Center, School of Physics and Astronomy, Shanghai Jiao Tong University, Shanghai 200240, China}
%
\date{\today}
\begin{abstract}
We present a theoretical framework called Lorentz quantum mechanics, where
the dynamics of a system is a complex Lorentz transformation in complex Minkowski space.
In contrast,  in usual quantum mechanics,  the dynamics is the unitary transformation in Hilbert space.
In our Lorentz quantum mechanics, there exist three types of states, space-like, light-like, and time-like.
Fundamental aspects are explored in parallel to the usual quantum mechanics, such as
matrix form of a Lorentz transformation, construction of  Pauli-like matrices for spinors.
We also investigate the adiabatic evolution in this mechanics, as well as the associated Berry curvature and Chern number. Three typical physical systems, where this Lorentz quantum dynamics
can arise,  are presented. They are  one dimensional fermion gas,
Bose-Einstein condensate (or superfluid), and one dimensional antiferromagnet.
\end{abstract}

\maketitle
\end{CJK}

\section{Introduction}

At the core of theoretical physics, two forms of vector transformations are of fundamental importance: the unitary transformation and the Lorentz transformation. The former, usually representing rotation of a real vector in space, preserves the modulus of the vector. In contrast, the latter, associated with the relativistic boost of a real vector in space-time, preserves the interval. In the context of quantum mechanics based on the Schr\"odinger equation, unitarity is an essential requirement for transformations of space, time, and spin, such that the modulus of a state vector in the Hilbert space -- representing the total probability of finding the particle -- is ensured invariant under these transformations. Instead, in this paper we address the quantum mechanics building on Lorentz transformations of complex vectors, where the temporal evolution and representation transformations conserve the interval. As we show below, such Lorentz quantum mechanics describes, and allow new insights into, the dynamical behavior of bosonic Bogoliubov quasiparticles.

We develop and study the Lorentz quantum mechanics basing on the Bogoliubov
equation~\cite{Bogoliubov,njp} for a $(1,1)$-type spinor, the simplest Lorentz spinor, with extensions to multi-mode spinors. In particular, we construct the matrix representing the Lorentz transformation of complex vectors, and the Lorentz counterpart of the standard Pauli matrices. Based on it, we explore in which ways the Lorentz quantum mechanics are similar to, and different from, the conventional quantum mechanics. We show that there exist many close analogies between the two, which allow extensions of, for example, the familiar adiabatic theorem and the concept of Berry phase to the context of Lorentz quantum mechanics. However, Lorentz time evolution can result in important modifications such as in the Berry connection \cite{ZhangNiu}.


We show that Lorentz spinors can generically arise in a variety of physical systems containing bosonic Bogoliubov quasiparticles. Specifically, we illustrate our study of the Lorentz quantum mechanics by investigating the spin wave excitations in a one dimensional (1D) antiferromagnetic system, the phonon excitations on top of a vortex in the Bose-Einstein condensate (BEC), and a 1D fermion gas at low temperatures. We note that an experimental proposal to observe the Berry phase effect on the dynamics of quasiparticles in a BEC with a vortex has been reported \cite{ZhangNiu}. Thus our present work  not only provide theoretically new insights into the dynamical properties of quasiparticles, but also allow feasible realization using the present experimental techniques with ultracold quantum gases.

The bosonic Bogoliubov equation has an identical form of the Schr\"odinger equation but governed by a non-Hermitian operator. In fact, the non-Hermitian Hamiltonian has been extensively studied in the context of PT-symmetric quantum mechanics, where the spectrum (eigenvalue) of non-Hermitian operator is proved to be real~\cite{Bender}. The PT-symmetric structure has found extensively applications in phonon-laser (coupled-resonator) system, where giant nonlinearity arises in the vicinity of phase transition between PT-symmetric phase and broken-PT phase, resulting in enhanced mechanical sensitivity~\cite{J1}, optical intensity~\cite{J2}, controllable chaos~\cite{J3} and optomechanically-induced transparency~\cite{J4}, as well as the phonon-rachet effect~\cite{J5}. The geometric phase of PT-symmetric quantum mechanics~\cite{Gong1} and the stability of driving non-Hermitian system has also been studied~\cite{Gong2}. The bosonic Bogoliubov operator studied here stands for a class of generalized PT symmetric Hamiltonian~\cite{wang}, or more precisely, the anti-PT Hamiltonian~\cite{XiaoNP}, which
can be realized experimentally by making use of refractive indices in optical settings~\cite{XiaoNP,GOP}.

\section{basic structures of Lorentz quantum mechanics}
\label{basic}
The Lorentz quantum mechanics is described by the following dynamical equation
\begin{equation} \label{evolution2}
{\text i}\hbar\frac{d}{dt}\left(\begin{array}{c}a_1(t)\\a_2(t) \\ \vdots \\a_{m+n}(t)\end{array} \right)=\sigma_{m,n} H \left(\begin{array}{c}a_1(t)\\a_2(t) \\ \vdots \\a_{m+n}(t)\end{array} \right)\,,
\end{equation}
where $H=H^\dag$ is a Hermitian matrix while $\sigma_{m,n}$ is given by
\begin{equation}
\sigma_{m,n}= {\text {diag}} \{ \underbrace{1,1,\ldots 1}_{m}, \underbrace{-1,-1,\ldots -1}_{n}  \}.
\end{equation}
Equations of this type are usually called Bogoliubov-de Gennes (BdG) equations and are obeyed
by bosonic quasi-particles in many different physical system (see Sec.~\ref{examples}).
For simplicity, we use the case $\sigma_{1,1}$ to explore the basic structures of the Lorentz
quantum mechanics as generalization to $\sigma_{m,n}$ is straightforward.

The BdG equation for spinor (1,1) is
\begin{equation} \label{evolution}
{\text i}\hbar\frac{d}{dt}\left(\begin{array}{c}a(t) \\b(t)\end{array} \right)=\sigma_{1,1} H \left(\begin{array}{c}a(t) \\b(t)\end{array} \right).
\end{equation}
Here $a(t)$ and $b(t)$ are the standard Bogoliubov amplitudes, $H=H^\dag$ is a Hermitian matrix, and $\sigma_{1,1}=\sigma_z$ is the familiar Pauli matrix in the $z$ direction, i.e.
\begin{equation}
\sigma_{1,1}= {\text {diag}} \{1,-1\} =\left(\begin{array}{cc}1&0\\ 0&-1\end{array} \right).
\end{equation}
The $\sigma_{1,1}H$ as the generator of the dynamics for spinor $(1,1)$
is an analogue of the Hamiltonian in the Schr\"odinger picture. Different from the Hamiltonian, though, $\sigma_{1,1} H$ is not Hermitian.

Note that the BdG equation stands for a special class of PT-symmetric quantum mechanics~\cite{wang,Bender2,Bender3,Bender4,Bender5}. The general form of two-mode PT-symmetric Hamiltonian has been written as~\cite{wang,Bender2,Bender3,Bender4,Bender5}
\begin{equation} \label{generalPT}
H_{PT}=\left(\begin{array}{cc} \epsilon+\gamma\cos\theta-{\text {i}}\mu\sin\theta & (\gamma\sin\theta+{\text {i}}\mu\cos\theta+\nu)e^{-{\text {i}}\varphi} \\
(\gamma\sin\theta+{\text {i}}\mu\cos\theta-\nu)e^{{\text {i}}\varphi} & \epsilon-\gamma\cos\theta+{\text {i}}\mu\sin\theta
 \end{array}\right),
\end{equation}
where $\epsilon$, $\mu$, $\nu$, $\gamma$, $\theta$ and $\varphi$ are real parameters.
In fact, both Hermitian Hamiltonian $H$ and BdG Hamiltonian $\sigma_{1,1}H$ constitute the subsets of (belong to) the PT-symmetric Hamiltonian (\ref{generalPT}).
It follows from (\ref{generalPT}) that the two-mode BdG Hamiltonian $\sigma_{1,1} H$ recovers from $H_{PT}$ when $\epsilon=0$, $\theta=2N\pi$ and $\varphi=2N\pi$; while the Hermitian Hamiltonian $H$ recovers when $\mu=\nu=0$.

For a Hermitian Hamiltonian $H$, we know that there are two eigenvectors, denoted $|1\rangle=(u,v)^T$ and $|2\rangle=(v^*,-u^*)^T$, where the normalized convention $|u|^2+|v|^2=1$ is usually employed. The orthogonality condition and completeness condition can be written as $\langle1|2\rangle=0$ and $|1\rangle\langle1|+|2\rangle\langle2|=1$, respectively. For an initial state $|\psi_0\rangle=(a(0),b(0))^T$, provided the dynamics is determined by the Schr\"odinger equation governed by the Hermitian Hamiltonian $H$, i.e.,
\begin{equation}
{\text i}\hbar\frac{d}{dt}\left(\begin{array}{c}a(t) \\b(t)\end{array} \right)=H \left(\begin{array}{c}a(t) \\b(t)\end{array} \right),
\end{equation}
it can be proven that the during the evolution of the wavefunction, the norm $|a(t)|^2+|b(t)|^2$ is conserved, i.e., the temporal evolution constitutes a unitary transformation. The representation transformation in unitary quantum mechanics must be a unitary transformation too.

However, as we shall see later, if the dynamics of a two-mode wavefunction is determined by the BdG equation (\ref{evolution}), the evolution definitely constitutes a complex Lorentz transformation in complex Minkowski space. The representation will also be associated with Lorentz transformations.

\subsection{Complex Lorentz transformation and complex Minkowski space}

Suppose the wavefunction's dynamics is governed by the BdG equation (\ref{evolution}), then for an arbitrary initial state $|\psi(0)\rangle=[a(0),b(0)]^T$, the wavefunction $|\psi(t)\rangle=[a(t),b(t)]^T$ at times $t>0$ can be solved formally from Eq. (\ref{evolution}) as
\begin{equation}\label{Eq:psit}
|\psi(t)\rangle=\mathcal{U}(t,0)|\psi(0)\rangle.
\end{equation}
Here $\mathcal{U}(t,0)$ is the evolution operator defined by
\begin{equation}\label{eq:U}
\mathcal{U}(t,0)=e^{-{\text i}\sigma_{1,1}Ht/\hbar}.
\end{equation}
The goal of this section is to show that the operator $\mathcal{U}(t,0)$ defined in Eq. (\ref{eq:U}) generates a \textit{complex Lorentz} - instead of a unitary - evolution of $|\psi(t)\rangle$.  In particular, defining the interval for a Lorentz spinor
\begin{equation}
\iv((a,b)^T)=(a^*,b^*)\sigma_{1,1}(a,b)^T=|a|^2-|b|^2,
\end{equation}
we prove below that the interval is conserved under the evolution generated by $\mathcal{U}(t,0)$, i.e.
\begin{equation}\label{eq:interval}
|a(t)|^2-|b(t)|^2=|a(0)|^2-|b(0)|^2.
\end{equation}

For above purpose, we first establish the following relation,
\begin{equation} \label{Lrelation}
\mathcal{U}^\dag\sigma_{1,1}\mathcal{U}=\sigma_{1,1}.
\end{equation}
Expanding $\sigma_{1,1}\mathcal{U}$ and $(\mathcal{U}^\dag)^{-1}\sigma_{1,1}$ in Taylor series, and noting $\sigma_{1,1}\sigma_{1,1}=1$, the $n$th term in the expansions of both $\sigma_{1,1}\mathcal{U}$ and $(\mathcal{U}^\dag)^{-1}\sigma_{1,1}$ are of the form
\begin{equation}
\frac{1}{n!}(-\frac{{\text i}}{\hbar})^n t^n H \underbrace{\sigma_{1,1}H\sigma_{1,1}H\ldots\sigma_{1,1}H}_{n-1 \,(\sigma_{1,1}H){\text s}}.
\end{equation}
This readily gives
\begin{equation}
\sigma_{1,1}\mathcal{U}=(\mathcal{U}^\dag)^{-1}\sigma_{1,1},
\end{equation}
from which Eq. (\ref{Lrelation}) ensues. Hence, by virtue of Eq. (\ref{Lrelation}), we obtain
\begin{equation} \label{L1}
\langle\psi(t)|\sigma_{1,1}|\psi(t)\rangle=\langle\psi(0)|\sigma_{1,1}|\psi(0)\rangle,
\end{equation}
and thus Eq.~(\ref{eq:interval}). In fact, the normalization of Lorentz-kind Eq.~(\ref{eq:interval}) has been extensively demonstrated in non-Hermitian quantum mechanics (see~\cite{Gong2} for an example).

Why can we refer to the equation (\ref{eq:interval}) as an analogue of Lorentz transformation? Since what we are focusing is the two-mode wavefunction, we can demonstrate this by the two dimensional space-time spanned by $(x,t)$. In special relativity, the interval $x^2-t^2$ (in natural units $c=1$) for a given inertial frame keeps a constant after the Lorentz boost to any other inertial frame. Here because $x$ and $t$ are both real numbers, $x^2-t^2=|x|^2-|t|^2$. The vector $(x,t)$ are called space-like, light-like and time-like as $x^2-t^2>0$, $x^2-t^2=0$ and $x^2-t^2<0$, respectively.

For the current two-mode wavefunction $(a,b)^T$, we can map the first component $a$ as $x$ and the second one $b$ as $t$. Thus the interval-like quantity $|a|^2-|b|^2$ can be accordingly defined. Because $a$ and $b$ may be complex numbers, the notion of modulus is necessary to define the interval. Since we have proven that, during the evolution determined by BdG equation, Eq.~(\ref{eq:interval}) holds, we can call this evolution as the Lorentz-like evolution, or complex Lorentz evolution. In analogy with the real Lorentz transformation, we consider that $(a,b)^T$ is space-like, light-like and time-like as ${\text {In}}((a,b)^T)=|a|^2-|b|^2>0$, ${\text {In}}((a,b)^T)=|a|^2-|b|^2=0$ and ${\text {In}}((a,b)^T)=|a|^2-|b|^2<0$, respectively.

While Eq.~(\ref{eq:interval}) formally resembles the conventional Lorentz evolution (transformation) in special relativity, there are delicate differences: (i) in contrast to the conventional Lorentz
transformation where only real numbers (space-time coordinate) are involved, here we are dealing with a complex vector specified by \textit{complex} numbers, the interval of which requires the notion of modulus (in this sense, we shall refer to the space where these complex vectors reside as the {\it complex Minkowski space}); (ii) unlike the real Minkowski space where $x(t)$ must be a space-like (time-like) component, here a freedom is left as we define the space-like axis and time-like one, i.e., we can either define $a(b)$ as the space-like (time-like) component or time-like (space-like) component. Thus whether a wavefunction $(a,b)^T$ is space-like or time-like is totally determined by how we define the space-like and time-like component. However, this does not constitute a problem as we can always fix our convention once the definition is determined.

We thus conclude that the evolution generated by $\mathcal{U}(t,0)$ conserves the interval [see Eq. (\ref{eq:interval})], and therefore, represents a \textit{complex Lorentz} evolution.

\subsection{Eigen-energies and eigenstates}

Although the $\sigma_{1,1} H$ is not Hermitian,  under certain conditions, it can admit real eigenvalues - which are relevant for physical processes. We write $\sigma_{1,1} H$ in terms
of three basic matrices as (dropping the term involving the identity matrix)
\begin{equation} \label{decomposition}
\sigma_{1,1}H=m_1\left(\begin{array}{cc}0&1 \\ -1&0\end{array} \right)+m_2\left(\begin{array}{cc}0&{\text i} \\ {\text i}&0\end{array} \right)+m_3\left(\begin{array}{cc}1&0 \\ 0&-1\end{array} \right),
\end{equation}
where the parameters $m_i$ ($i=1,2,3$) are real. The eigen-energies  are  the
roots of the following equation
\begin{equation}
\label{m3}
m_3^2-(m_1^2+m_2^2)=E^2\,.
\end{equation}
It is clear that the eigenvalues are real provided the condition
\begin{equation} \label{first-condition}
m_3^2\geq m_1^2+m_2^2
\end{equation}
is satisfied. In this work, we shall restrict ourselves to this physically relevant
regime of real-eigenvalues in the parameter domain specified by $(m_1,m_2,m_3)$,
and we denote the two real eigenvalues as $E_1$ and $E_2$, with the corresponding
eigenstates labeled as $|1\rangle$ and $|2\rangle$, respectively.

\begin{figure}
\includegraphics[width=0.90\linewidth]{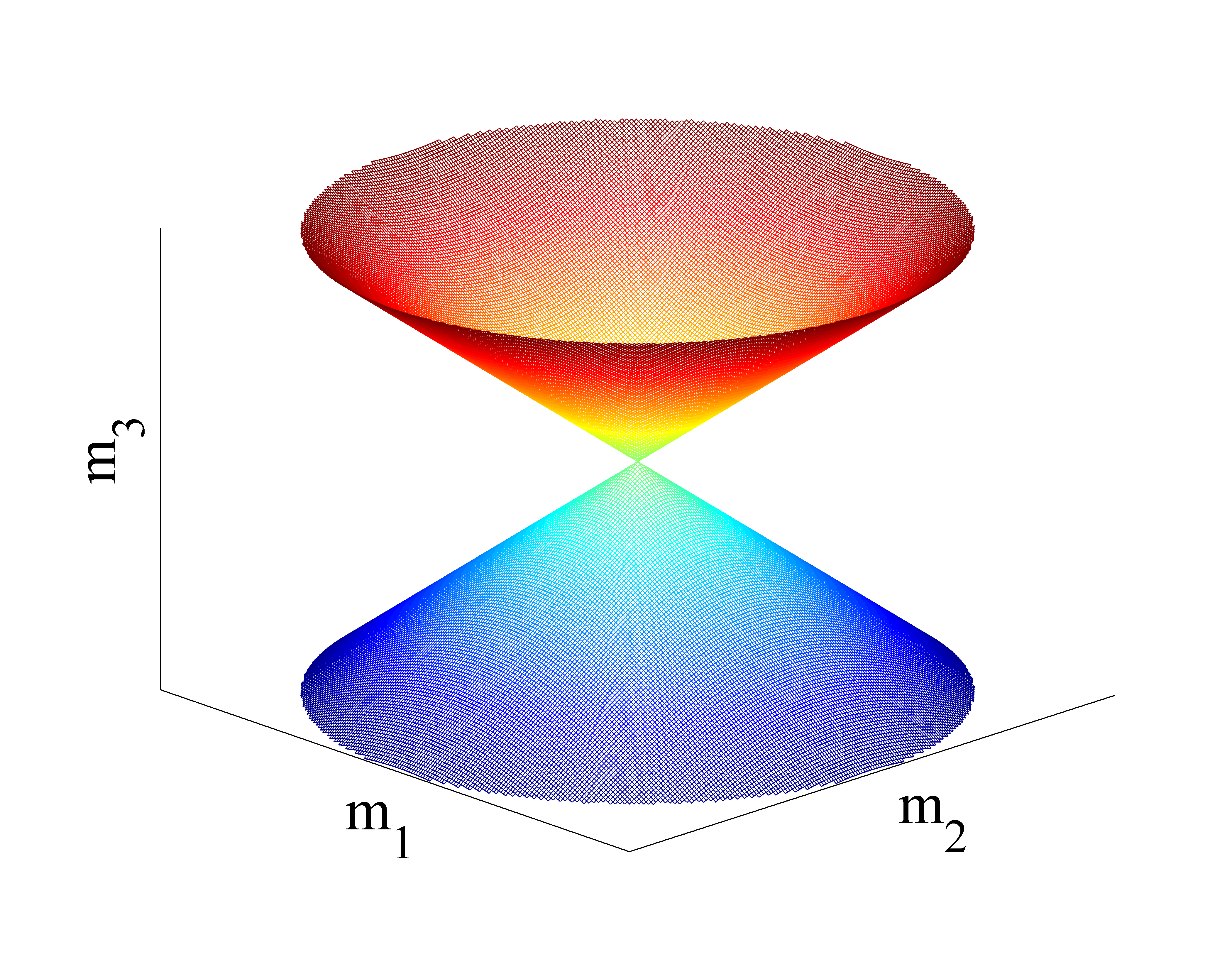}
\caption{(color online) The degeneracy regime of Lorentz spinor parameterized by $m_1$, $m_2$ and $m_3$ as in Eq.~(\ref{decomposition}) forms the surface of a cone. As will be discussed in
Sec.~\ref{berry}, the charge (monopole) for Berry curvature (monopole) is at the tip of the cone rather than distributing over the whole degeneracy cone.}
\label{Fig:1}
\end{figure}

\begin{figure}[t]
\includegraphics[width=0.95\linewidth]{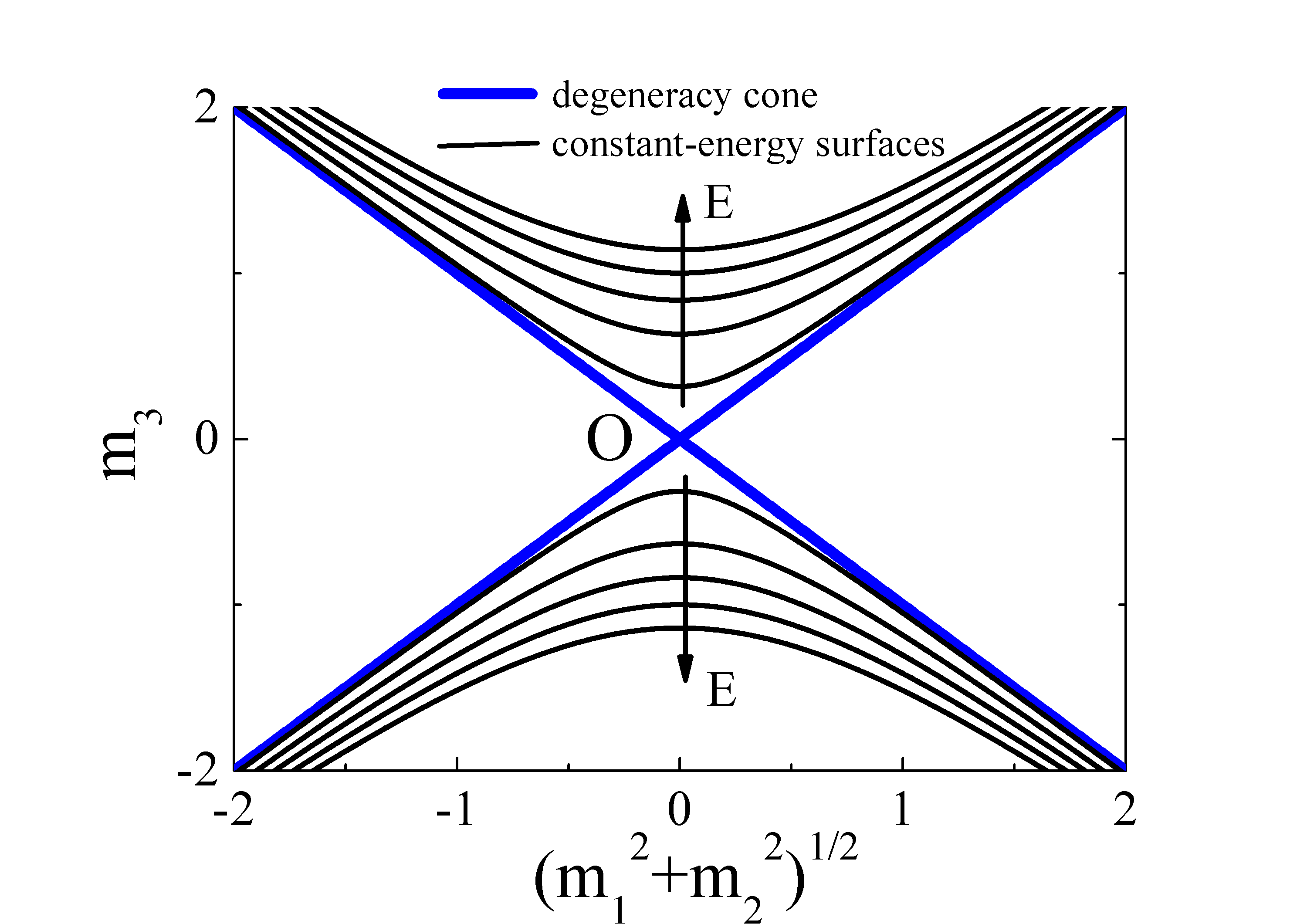}
\caption{(color online) Illustration of the constant-energy surfaces of BdG equation parameterized by $m_1$, $m_2$ and $m_3$. The arrows indicate the directions of increasing (decreasing) of energy for state $|1\rangle$ ($|2\rangle$). On the cone's surface, the two eigenstates are degenerate. Because the surfaces assume the axial symmetry about the $m_3$ axis, the two dimensional plot is depicted for clarity.}
\label{Fig:2}
\end{figure}

Two facts are clear from Eq.(\ref{m3}): ({\it i}) in the parameter space $(m_1,m_2,m_3)$, the two eigenstates $|1\rangle$ and $|2\rangle$ exhibit degeneracies on a circular cone (see Fig. \ref{Fig:1}), which resembles the light-cone in special relativity. This is in marked contrast to a unitary spinor,
where the degeneracy occurs only at an isolated point; ({\it ii}) unlike a unitary spinor where the constant-energy surfaces are  elliptic surfaces, both eigenstates of $\sigma_{1,1}H$ display hyperbolic constant-energy surfaces (see Fig. \ref{Fig:2}).

We now describe the basic properties of the eigenstates associated with the operator $\sigma_{1,1}H$. They are solutions to the following eign-equations
\begin{eqnarray} \label{eigenequation}
\sigma_{1,1} H|1\rangle&=&E_1|1\rangle, \\ \label{eigenequation2}
\sigma_{1,1} H|2\rangle&=&E_2|2\rangle.
\end{eqnarray}
Keeping in mind that only real eigenvalues are considered, for $E_1\neq E_2^*=E_2$, we have
\begin{equation} \label{orthoF}
\langle 2 |\sigma_{1,1}|1\rangle=0.
\end{equation}
It can be checked that the two eigenstates of $\sigma_{1,1}H$ can always be specifically expressed as \begin{equation}  \label{eigen-vec}
|1\rangle=\left(\begin{array}{c}u \\v\end{array} \right); \quad
|2\rangle=\left(\begin{array}{c}v^* \\u^*\end{array} \right).
\end{equation}
This means that if $\ket{1}$ is space-like then $\ket{2}$ is time-like or vice versa.

In the energy representation defined in terms of $|1\rangle$ and $|2\rangle$, a time-evolved state $|\psi(t)\rangle=[a(t),b(t)]^T$ [see Eq. (\ref{evolution})] can be written as
\begin{equation} \label{evo-form}
|\psi(t)\rangle=c_1|1\rangle e^{-{\text i}E_1t}+c_2
|2\rangle e^{-{\text i}E_2t}.
\end{equation}

In transforming $|\psi(t)\rangle$ from the Bogoliubov representation to the energy representation, the interval of the Lorentz spinor is preserved, i.e. it is a complex Lorentz transformation. To see this, using Eq. (\ref{orthoF}), we find
\begin{eqnarray} \label{Interval1}
{\text {In}}(|\psi(t)\rangle)&=&\langle\psi(t)|\sigma_{1,1}|\psi(t)\rangle \\
&=&|c_1|^2 \langle 1|\sigma_{1,1}|1\rangle+|c_2|^2 \langle 2|\sigma_{1,1}|2\rangle.
\end{eqnarray}
By further assuming a gauge for Lorentz-like normalization, i.e.,
\begin{eqnarray} \label{normalization} \nonumber
{\text {In}}(|1\rangle)&=&\langle1|\sigma_{1,1}|1\rangle=1, \\  {\text {In}}(|2\rangle)&=&\langle2|\sigma_{1,1}|2\rangle=-1,
\end{eqnarray}
we obtain from (\ref{Interval1}) that
\begin{equation}
\label{con2}
{\text {In}}(|\psi(t)\rangle)=|a|^2-|b|^2=|c_1|^2-|c_2|^2,
\end{equation}
meaning the interval is conserved for the above representation transformation.

The normalization condition $|u|^2-|v|^2=1$ is different from  the eigenstates of a conventional unitary spinor. In fact, if one naively enforce the unitary gauge on Eq. (\ref{eigen-vec}), say, $|u|^2+|v|^2=1$, \textit{unphysical} consequences would ensue: The time-evolved wavefunction in the original Bogoliubov representation [$|\psi\rangle=(a,b)^T$] could not maintain its ordinary amplitude, such that $|a(t)|^2+|b(t)|^2\neq1$ for $t>0$, and, in particular, the amplitude in different representation would take different value, e.g., $|c_1|^2+|c_2|^2\neq|a(t)|^2+|b(t)|^2$, which can be easily inferred from Eq.~(\ref{evo-form}).

In general, when $\sigma_{1,1}H$ takes the form (\ref{decomposition}) with $m_3=0$, it exhibits two light-like eigenvectors; whereas, when $m_3\neq0$, there are one space-like and one time-like eigenvectors. Thus, in the physically relevant regime $m_3^2\geq m_1^2+m_2^2$ as considered here, we find $|1\rangle$ is space-like and $|2\rangle$ time-like. As a result,  a light-like vector can be formed from a superposition of two eigenvectors with equal weight, i.e., $|c_1|^2-|c_2|^2=|a|^2-|b|^2=0$.

\subsection{Representation transformation and physical meaning of the wavefunction}

In the usual quantum mechanics, the change from one representation
to another (or from one basis to another) is given by a unitary matrix.
As discussed above,  the change  from the Bogoliubov
representation to the energy representation [see Eq.~(\ref{evo-form}) and Eq.~(\ref{con2})] is
facilitated by a Lorentz transformation. This motivates us to introduce a complex Lorentz operator $\mathfrak{L}$ acting on the Lorentz $(1,1)$-spinor, defined by
\begin{equation}\label{eq:L}
\mathfrak{L}=\left(\begin{array}{cc}x&y^* \\ y&x^*\end{array} \right),
\end{equation}
where $|x|^2-|y|^2=1$, with the corresponding inverse Lorentz matrix being
\begin{equation}
\mathfrak{L}^{-1}=\left(\begin{array}{cc}x^*&-y^* \\ -y&x\end{array} \right).
\end{equation}
Using the identity $\mathfrak{L}^\dag \sigma_{1,1}\mathfrak{L}=\sigma_{1,1}$, it is readily to see that both $\mathfrak{L}$ and $\mathfrak{L}^{-1}$ are Lorentz matrices.

For an arbitrary Lorentz matrix, we have,
\begin{equation}
{\text {In}}(|\psi\rangle)={\text {In}}(\mathfrak{L}|\psi\rangle),
\end{equation}
meaning the interval is preserved. Under a Lorentz transformation, an arbitrary physical operator $K$ transforms as
\begin{equation} \label{L-T-M}
K\rightarrow K'=\mathfrak{L}K\mathfrak{L}^{-1},
\end{equation}
while the corresponding eigenvalues stay unchanged. Note that, since $\mathfrak{L}$ is no longer a unitary matrix, we have $\mathfrak{L}^{-1}\neq\mathfrak{L}^{\dag}$.

To illustrate the above constructions, consider the transformation from the Bogoliubov to the energy representation as described earlier. In this case, the eigenstates $|1\rangle$ and $|2\rangle$ transform as
\begin{eqnarray}
|1\rangle&=&\left(\begin{array}{c}u \\v\end{array} \right)\rightarrow  \mathfrak{L}_{\text {B}} |1\rangle=\left(\begin{array}{c}1 \\0\end{array} \right) \\
|2\rangle&=&\left(\begin{array}{c}v^* \\u^*\end{array} \right)\rightarrow \mathfrak{L}_{\text {B}}|2\rangle=\left(\begin{array}{c}0 \\1\end{array} \right),
\end{eqnarray}
where the matrix $\mathfrak{L}_{\text {B}}$ is shown in Eq.~(\ref{eq:L}), with $x=u^*$ and $y=-v$, i.e.,
\begin{equation}\label{eq:LB}
\mathfrak{L}_{\text {B}}=\left(\begin{array}{cc}u^*&-v^* \\ -v&u\end{array} \right).
\end{equation}
Because now $|x|^2-|y|^2=|u|^2-|v|^2=1$, $\mathfrak{L}_{\text {B}}$, as shown in Sec.~IIC, must be a Lorentz matrix. Obviously, as we have proven, the interval must be conserved, i.e., $|u|^2-|v|^2=1^2-0^2=1$, $|v^*|^2-|u^*|^2=0^2-1^2=-1$. In addition, the Bogoliubov operator transforms as,
\begin{equation}
\sigma_{1,1}H\rightarrow \mathfrak{L}_{\text {B}}\sigma_{1,1}H \mathfrak{L}_{\text {B}}^{-1}=\left(\begin{array}{cc}E_1&0 \\0&E_2\end{array} \right).
\end{equation}
This special Lorentz transformation from the original representation to the energy representation is in fact equivalent to the Bosonic Bogoliubov transformation. This has been studied in the Swanson Hamiltonian~\cite{Swanson1}, where it is found that the energy eigenstates can be constructed from the algebra and states of the harmonic oscillator and transition probabilities governed by the non-Hermitian Swanson Hamiltonian are shown to be manifestly unitary. For a time-dependent Swanson Hamiltonian~\cite{Swanson2}, time-dependent Dyson and quasi-Hermiticity relation is demonstrated clearly.

In light of the conservation of interval - rather than norm - of the state vector under transformations, a question immediately arises as to whether, or to what extent, the wavefunction in the context of Lorentz quantum mechanics still affords the physical interpretation as the probability wave? Indeed, in the energy representation, see Eq.~(\ref{evo-form}), it is clear that $|c_{1(2)}|^2$, with $|c_1|^2+|c_2|^2=1$, can be interpreted as the probability of finding the spinor in the eigenstate $|1(2)\rangle$, i.e., a wavefunction $c_1|1\rangle+c_2|2\rangle$ still describes a probability wave. However, in the Bogoliubov representation, the interpretation of a wavefunction as the probability wave is no longer physically meaningful. For example, consider the eigenstate $|1\rangle=(u,v)^T$, which is usually generated from creating a pair of Bogoliubov quasiparticles in the ground state of the system. Yet, $|u|^2$ and $|v|^2$ cannot represent the probabilities in the Bogoliubov basis: the Bogoliubov basis is not a set of orthonormal basis (see Sec.~IV for concrete examples), and therefore, instead of $|u|^2+|v|^2=1$, the convention $|u|^2-|v|^2=1$ must be taken.

\subsection{Completeness of eigenvectors}
Based on Eq. (\ref{eigen-vec}) [see also Eqs. (\ref{orthoF}) and (\ref{normalization})], the completeness of eigenvectors in the energy representation now takes a different form compared to the unitary case, reading
\begin{equation} \label{completeness1}
\bigsqcup_{j} |j\rangle\langle j| \sigma_{1,1}=1,
\end{equation}
or, equivalently,
\begin{equation} \label{completeness2}
\sigma_{1,1} \bigsqcup_{j}  |j\rangle\langle j|=1.
\end{equation}
Here, the notation $\bigsqcup_j$ [for $(1+1)$-mode] is defined by
\begin{equation}
 \bigsqcup_{j}  |j\rangle\langle j|=|1\rangle\langle 1|-|2\rangle\langle 2|.
\end{equation}
It can be found easily that, ensured by the property of Lorentz matrix $\mathfrak{L}^\dag \sigma_{1,1}\mathfrak{L}=\sigma_{1,1}$, the completeness expression (\ref{completeness1}) (or (\ref{completeness2})) remains in any other representation.

\subsection{Analogue of Pauli Matrices}
In analogy with the conventional spinor that is acted by the basic operators known as Pauli matrices, it is natural to ask, for the Lorentz spinor, if similar matrices can be constructed. Such analogue of the Pauli matrices, denoted by $\tau_i$ ($i=1,2,3$), is required to fulfill the following conditions: (i) any operator $\sigma_{1,1} H$, when written in terms of $\tau_i$ (dropping the term involving identity matrix), i.e.,
\begin{equation} \label{se-decomposition}
\sigma_{1,1} H=n_1\tau_1+n_2\tau_2+n_3\tau_3,
\end{equation}
must have real-number components $n_i$; (ii) the matrices $\tau_i$ ($i=1,2,3$) should have the same real eigenvalues, say, $\pm1$, and can transform into each other via Lorentz transformation [see Eq. (\ref{L-T-M})].

Based on (i) and (ii), we see that the matrices as appeared in Eq.~(\ref{decomposition}) do not represent the analogue of the Pauli matrix for the Lorentz spinor: while they
satisfy the requirement (i), the condition (ii) is violated. Instead, we consider following constructions:
\begin{equation}\label{eq:pauli}
\tau_1=\left(\begin{array}{cc}\sqrt{2}&1 \\ -1&-\sqrt{2}\end{array} \right), \medspace \tau_2=\left(\begin{array}{cc}\sqrt{2}&{\text i} \\ {\text i}&-\sqrt{2}\end{array} \right), \medspace \tau_3=\left(\begin{array}{cc}1&0 \\ 0&-1\end{array} \right).
\end{equation}
It is easy to check that $\tau_i$ in Eq. (\ref{eq:pauli}) satisfy both requirements (i) and (ii). In particular, the transformation between $\tau_1$ and $\tau_3$ is explicitly found to be
\begin{equation}
\tau_1=\mathfrak{L}\tau_3\mathfrak{L}^{-1},
\end{equation}
where $\mathfrak{L}$ is of the form (\ref{eq:L}) with $x=\frac{\sqrt{2}+1}{2}{\text i}$ and $y=-\frac{\sqrt{2}-1}{2}{\text i}$, and that between $\tau_2$ and $\tau_3$ is given by
\begin{equation}
\tau_2=\mathfrak{L}\tau_3\mathfrak{L}^{-1},
\end{equation}
for  $\mathfrak{L}$ with $x=\frac{\sqrt{2}+1}{2}e^{-{\text i}\frac{\pi}{4}}$ and $y=-\frac{\sqrt{2}-1}{2}e^{{\text i}\frac{\pi}{4}}$.

\subsection{Heisenberg picture}

The current Lorentz evolution is in fact defined in the analogue of Schr\"odinger picture (denoted by subscript $s$), i.e., any physical operator keeps constant while the wavefunction undergoes Lorentz evolution. In analogy with the conventional spinor, the Lorentz quantum mechanics can also be expressed in the analogue of Heisenberg picture (denoted by subscript $h$). The relations of an operator $\mathcal{O}$ and the state $|\psi\rangle$ between the two pictures are,
\begin{eqnarray}
\mathcal{O}(t)_h&=&e^{i\sigma_{1,1}Ht}\mathcal{O}_s e^{-i\sigma_{1,1}Ht}, \\
|\psi\rangle_h&=&e^{i\sigma_{1,1}Ht} |\psi(t)\rangle_s,
\end{eqnarray}
where $|\psi\rangle_h$ keeps constant but $\mathcal{O}(t)_h$ satisfies the analogue of Heisenberg equation,
\begin{equation}
i\hbar \frac{\partial \mathcal{O}(t)_h }{\partial t}=[\sigma_{1,1}H,\mathcal{O}(t)_h],
\end{equation}
with $[\sigma_{1,1}H,\mathcal{O}(t)_h]$ being the commutator between $\sigma_{1,1}H$ and $\mathcal{O}(t)_h$.

\subsection{Generalization to multi-mode}

In this section, we extend the above formulations for the $\sigma_{1,1}$ Lorentz spinor to the case of multi-mode spinor with $\sigma_{m,n}$. The operator $\sigma_{m,n}H$ has $m+n$ energy eigenstates, denoted by $|1\rangle$, $|2\rangle$, $\ldots$, $|m+n\rangle$.
Define the interval of a $(m+n)$-mode wavefunction $|\psi\rangle=(a_1,a_2,\ldots,a_{m+n})^T$ as,
\begin{equation}
{\text {In}}(|\psi\rangle)=\langle\psi|\sigma_{m,n}|\psi\rangle=\sum_{j=1}^{m} |a_j|^2-\sum_{j=m+1}^{m+n}  |a_j|^2.	
\end{equation}
It is easy to see that the intervals of the eigenstates are,
\begin{eqnarray}\label{intervm}
{\text {In}}(|j\rangle)&=&1 \quad {\text{for}} \quad j=1,2,\ldots m, \\ \nonumber
{\text {In}}(|j\rangle)&=&-1 \quad {\text{for}}\quad j=m+1,m+2,\ldots m+n.
\end{eqnarray}

In addition, the orthogonal condition for two non-degenerate eigenstates is derived as,
\begin{equation}\label{orthoFm}
\langle j|\sigma_{m,n}|k\rangle=0, \quad {\text{for}} \quad j\neq k,
\end{equation}
generalizing Eq. (\ref{orthoF}) for the $(1,1)$-mode. Using Eqs.  (\ref{intervm}) and (\ref{orthoFm}), the completeness of eigenvectors can be expressed as
\begin{equation} \label{completeness3}
\bigsqcup_{j} |j\rangle\langle j| \sigma_{m,n}=1,
\end{equation}
or, equivalently,
\begin{equation} \label{completeness4}
\sigma_{m,n} \bigsqcup_{j}  |j\rangle\langle j|=1,
\end{equation}
with the symbol $\bigsqcup_j$ for $(m,n)$-mode defined as
\begin{equation}
 \bigsqcup_{j}  |j\rangle\langle j|=\sum_{j=1}^m|j\rangle\langle j|-\sum_{j=m+1}^{m+n}|j\rangle\langle j|.
\end{equation}

\section{Adiabaticity and geometric phase}
\label{berry}

\subsection{Adiabatic theorem}

Consider a $(1,1)$-spinor described by the operator $\sigma_{1,1}H(\mathbf{R})$, which depends on a set of system's parameter $\mathbf{R}$. Suppose the spinor is initially in an eigenstate, say $|1\rangle$, before the parameter $\mathbf{R}$ undergoes a sufficiently slow variation, thus driving an adiabatic evolution for the Lorentz spinor.  The relevant matrix element capturing the slowly varying time-dependent perturbation can be evaluated as, by acting the gradient operator $\nabla\equiv\frac{\partial}{\partial \mathbf{R}}$ on the Eq. (\ref{eigenequation}) and using Eq. (\ref{eigenequation2}),
\begin{equation} \label{for-Berry-adia}
\langle 2|\sigma_{1,1} \nabla|1\rangle=\frac{\langle 2|\nabla H|1\rangle}{E_1-E_2^*}=\frac{\langle 2|\nabla H|1\rangle}{E_1-E_2}.
\end{equation}
Here, the last equality is ensured by the real eigenvalues in the considered parameter regimes, together with the condition $E_1\neq E_2$.

We see that the relation (\ref{for-Berry-adia}), except for an additional $\sigma_{1,1}$, is identical with that in unitary quantum mechanics \cite{BornFock}. This allows us to generalize the familiar adiabatic theorem to the context of Lorentz quantum mechanics: Starting from an initial eigenstate $|1(\mathbf{R})\rangle$ ($|2(\mathbf{R})\rangle$), the system will always be constrained in this instantaneous eigenstate so long as $\mathbf{R}$ is swept slowly enough in the parameter space. (A rigorous proof would be similar to that in the conventional quantum mechanics~\cite{BornFock,Zhang},
and therefore, here we shall leave out the detailed procedure.)

\subsection{Analogue of Berry phase}

In conventional quantum mechanics, it is well known that an eigen-energy state undergoing an adiabatic evolution will pick up a Berry phase~\cite{Berry}, when a slowly varying system parameter $\mathbf{R}$ realizes a loop in the parameter space. Here we show that in the context of Lorentz quantum mechanics, a Lorentz counterpart of the Berry phase will similarly arise.

The time evolution of an instantaneous eigenstate, which is parametrically dependent on $\mathbf{R}$, can be written as
\begin{equation} \label{ansatz}
|\psi\rangle=|m\rangle e^{-{\text i}\frac{\int E_m(\mathbf{R}) dt}{\hbar}}e^{i\beta},
\end{equation}
with $m=1,2$. Here, $-\int E_m(\mathbf{R})dt/\hbar$ denotes the dynamical phase and $\beta$ the geometric phase. Substituting Eq. (\ref{ansatz}) into Eq. (\ref{evolution}), we find
\begin{equation}\label{eq:beta1}
\frac{d\beta_1}{d\mathbf{R}}={\text i}\langle1|\sigma_{1,1} \frac{\partial}{\partial \mathbf{R}}|1\rangle;
\end{equation}
and
\begin{equation}\label{eq:beta2}
\frac{d\beta_2}{d\mathbf{R}}=-{\text i}\langle2|\sigma_{1,1} \frac{\partial}{\partial \mathbf{R}}|2\rangle.
\end{equation}
From Eqs, (\ref{eq:beta1}) and (\ref{eq:beta2}), we can readily read off the Berry connections as
\begin{eqnarray} \label{Berry-con1}
 \mathbf{A}_1={\text i}\langle 1|\sigma_{1,1} \nabla|1\rangle, \\  \label{Berry-con2}
 \quad \mathbf{A}_2=-{\text i}\langle 2|\sigma_{1,1} \nabla|2\rangle.
\end{eqnarray}

Equations (\ref{Berry-con1}) and (\ref{Berry-con2}) show that the Berry connection in the Lorentz quantum mechanics is modified from the conventional one, where the Berry connection is given by $i\langle m|\frac{\partial}{\partial \mathbf{R}}|m\rangle$. Will such modifications give rise to a different monopole structure for the Berry curvature? Or, will the monopole in the Lorentz mechanics still occur at the degeneracy point (where $E_1=E_2$)?  To address these questions, we now calculate the Berry curvature $\mathbf{B}=\nabla\times \mathbf{A}$. Without loss of generality, we take the eigenvector $|1\rangle$ for concrete calculations.

Our starting point is the identity $\langle 1 |\sigma_{1,1}|1\rangle=1$. By acting $\nabla$ on both sides, we obtain
\begin{equation}
\langle 1|\sigma_{1,1} \nabla |1\rangle+\langle 1|\sigma_{1,1} \nabla |1\rangle^*=0.
\end{equation}
This indicates that $\langle 1|\sigma_{1,1} \nabla |1\rangle$ is purely imaginary ($\mathbf{A}_1$ is real). Hence, $\mathbf{B}_1$ can be evaluated as,
\begin{equation} 
\mathbf{B}_1=\nabla\times \mathbf{A}_1
=-{\text I_m}\bigsqcup_j \langle \nabla1 |\sigma_{1,1}|j\rangle\langle j| \sigma_{1,1}\times\nabla |1\rangle\,, \label{B1}
\end{equation}
where ${\text I_m}$ represents the imaginary part.
In deriving  Eq. (\ref{B1}), we have used  the completeness relation (\ref{completeness1}) and the following relation
\begin{equation} \label{are}
\nabla\times(\mu \mathbf{b})=\nabla\mu\times\mathbf{b}+\mu\nabla\times\mathbf{b},
\end{equation}
valid for arbitrary scalar $\mu$ and vector $\mathbf{b}$.

According to Eq. (\ref{for-Berry-adia}), $\mathbf{B}_1$ in Eq. (\ref{B1}) is well defined provided $E_1\neq E_2$, such that the monopole is expected to be absent in this case. To rigorously establish this,  let us calculate the divergence of the Berry curvature, i.e. $\nabla \cdot \mathbf{B}_1$. Introducing an auxiliary operator,
\begin{equation}
\mathbf{F}=-{\text i}\sigma_{1,1}\bigsqcup_j|\nabla j\rangle\langle j|\sigma_{1,1},
\end{equation}
which is Hermitian, $\mathbf{F}=\mathbf{F}^\dag$, as ensured by the completeness relation (\ref{completeness1}), we have
\begin{eqnarray}  \label{F-re} \nonumber
\sigma_{1,1}|\nabla j\rangle&=&{\text i}\mathbf{F}|j\rangle \\ \nonumber
\nabla\times\mathbf{F}&=&-{\text i}\sigma_{1,1}\bigsqcup_j|\nabla j\rangle\times\langle \nabla j|\sigma_{1,1} \\ \nonumber
&=&-{\text i}\bigsqcup_j\mathbf{F}|j\rangle\times\langle j|\mathbf{F} \\
&=&-{\text i}\mathbf{F}\times\sigma_{1,1}\mathbf{F}.
\end{eqnarray}
In deriving above, we have used Eq. (\ref{are}).  Further noting that
\begin{equation}
{\text i}\langle j|\mathbf{F}|k\rangle=\bigsqcup_{j'}\langle j|\sigma_{1,1}|\nabla j'\rangle\langle j'|\sigma_{1,1}|k\rangle=\langle j|\sigma_{1,1}\nabla|k\rangle,
\end{equation}
the Berry curvature can be expressed in terms of $\mathbf{F}$ as
\begin{equation} 
\mathbf{B}_1=-{\text I_m} \bigsqcup_j \langle1|\mathbf{F}|j\rangle\times\langle j|\mathbf{F}|1\rangle
=-{\text I_m}\langle1|\mathbf{F}\times\sigma_{1,1}\mathbf{F}|1\rangle.
\end{equation}
Finally, by virtue of $\nabla\times\mathbf{F}$ in Eq.~(\ref{F-re}), we find
\begin{eqnarray} \nonumber
\nabla\cdot\mathbf{B}_1&=&-{\text I_m}[\langle\nabla1|\cdot(\mathbf{F}\times\sigma_{1,1}\mathbf{F})|1\rangle+
\langle1|(\mathbf{F}\times\sigma_{1,1}\mathbf{F})\cdot\nabla|1\rangle \\ \nonumber
&&+\langle1|\nabla\cdot(\mathbf{F}\times\sigma_{1,1}\mathbf{F})|1\rangle] \\ \nonumber
&=&-{\text I_m}[-{\text i}\langle1|\mathbf{F}\sigma_{1,1}\cdot(\mathbf{F}\times\sigma_{1,1}\mathbf{F})|1\rangle+{\text i}\langle1|(\mathbf{F}\times\sigma_{1,1}\mathbf{F})\cdot\sigma_{1,1}\mathbf{F}|1\rangle \\ \nonumber
&&+\langle1|(\nabla\times\mathbf{F})\cdot\sigma_{1,1}\mathbf{F}|1\rangle-\langle1|\mathbf{F}\sigma_{1,1}\cdot(\nabla\times\mathbf{F})|1\rangle ] \\ \nonumber
&=&0.
\end{eqnarray}
Therefore, as expected, the monopole in the Lorentz quantum mechanics can only appear in the degenerate regime where $\mathbf{B}_1$ diverges, similar as the conventional unitary quantum mechanics.

Next, searching for the monopole, we focus on the degeneracy regime in the parameter space defined by $(m_1,m_2,m_3)$, which, as shown in Fig.~1, forms a circular cone. There, imagine the path of $\mathbf{R}=(m_1,m_2,m_3)$ realizes a loop in the vicinity of the cone's surface. In this case, the instantaneous eigenstate, say, $|1(\mathbf{R})\rangle$, is expected to vary in a back-and-forth manner (dropping the overall phases including both the dynamical and Berry phase). This is because the instantaneous eigenstate, apart from an overall phase, is always the same along any straight line emanating from the origin. As a result, the integration of $\mathbf{A}_1$ along this loop vanishes, meaning there is no charge of the Berry curvature on the cone's surface, even though it is in the degeneracy regime.

We thus conclude that - just as in the case of unitary spinor - the charge, if exists, can only be distributed on the isolated points, i.e., the original monopole $O$, in $\mathbf{R}=(m_1,m_2,m_3)$ space. However, different from unitary spinor, the magnetic flux does not uniformly emanate from the monopole $O$ to the parameter space, instead, it emanates only to the region in the cone (more closer to the $m_3$ axis). In addition, even in this region, the magnetic flux is not uniformly distributed. Specifically, by evaluating the geometric phase along a loop perpendicular to the $m_3$ axis, we can find the distribution of the magnetic flux density per {\it solid angle} as a function of the angle $\theta$ from $m_3$ axis, i.e.,
\begin{equation} \label{density}
\rho=\mp \frac{(1+\tan^2\theta)^{\frac{3}{2}}}{2(1-\tan^2\theta)^{\frac{3}{2}}},
\end{equation}
with $-/+$ associated with the state $|1\rangle$ ($|2\rangle$). Note that the flux density is proportional to the Berry curvature, which acts as a magnetic field, whose magnitude according to Eq. (\ref{density}) increases when approaching the cone. Right on the surface of the cone, where $\theta\rightarrow\frac{\pi}{4}$, the magnetic field diverges. Outside the cone, on the other hand, the eigenvalue becomes complex such that the notion of adiabatic evolution and geometric phase become meaningless, i.e., there is no magnetic field emanating outside the cone from the monopole $O$. Again, due to the aforementioned fact that the instantaneous eigenstate (apart from an overall phase) remains the same along any straight line emanating from the origin, we expect all the magnetic field fluxes to be described by straight lines (see Fig.~3).

Alternatively, we can write $\sigma_{1,1} H$ in terms of the analogues of Pauli's matrices $\tau_i$  [see Eq.~(\ref{se-decomposition})], which is then mapped onto a vector $(n_1, n_2, n_3)$ in the parameter space. However, this equivalent kind of decomposition will not contribute anything but modify the slope of Berry curvature $\theta\rightarrow\theta'$ ($\tan(\theta)=1/C$, while $\tan(\theta')=1/(C-\sqrt{2})$, with $C$ being any constant).

\subsection{Chern number}

The Chern number - which reflects the total magnetic charge contained by the monopole on $O$ - can be calculated from Eq.~(\ref{density}) as,
\begin{equation} 
\mathcal{C}_n=\mp \infty,
\end{equation}
with $-/+$ for the state $|1\rangle$ ($|2\rangle$). Hence, the Lorentz spinor not only has distinct distribution of the magnetic flux compared to the unitary spinor, but also possesses unexpectedly the qualitatively different Chern number which is divergent.

\begin{figure}[t]
\includegraphics[width=0.95\linewidth]{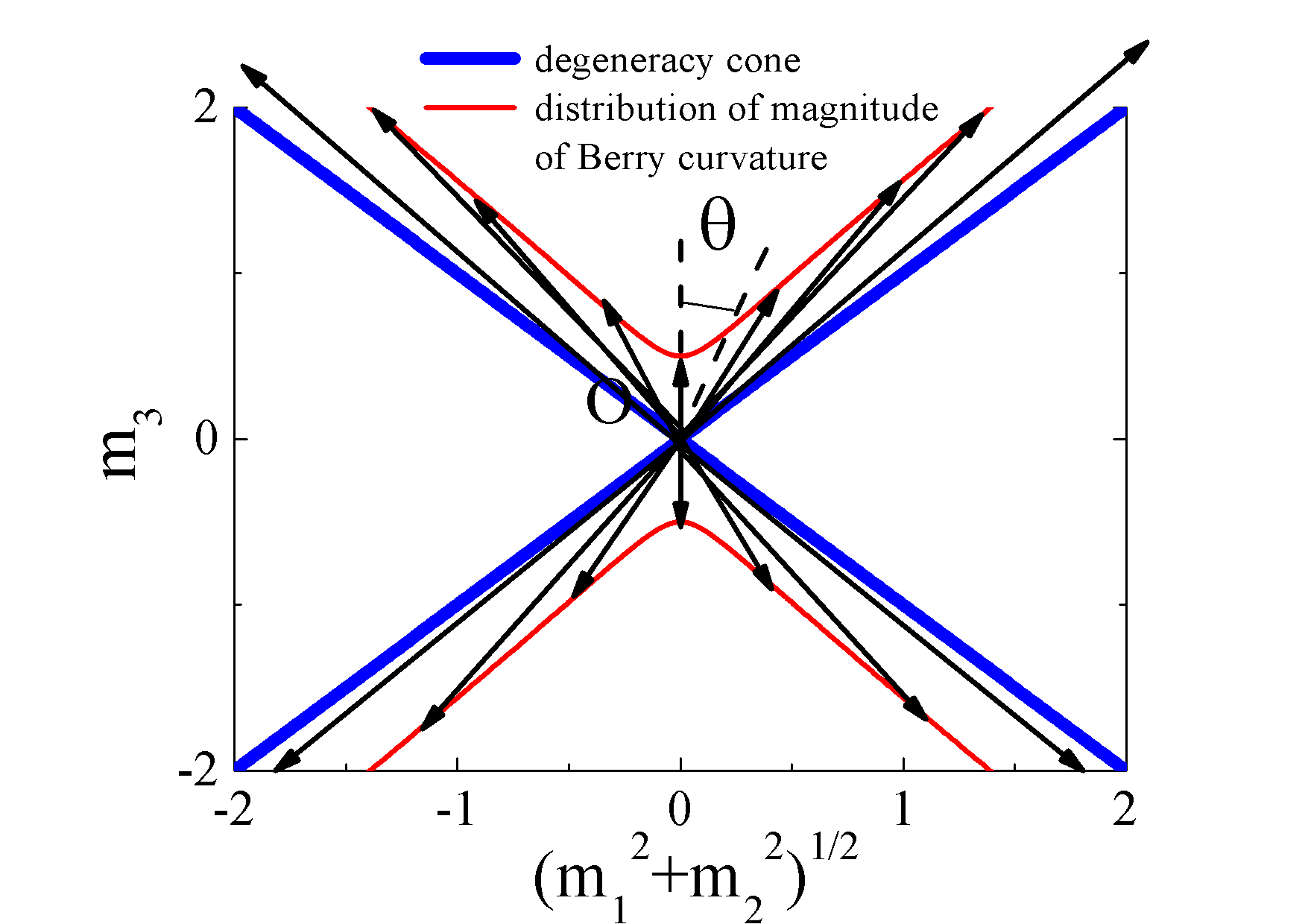}
\caption{(color online) Illustration of the analytic result given by Eq.~(\ref{density}) for the distribution of strength of Berry curvature (magnetic field) for instantaneous eigenstate $|2\rangle$. For the state $|1\rangle$, everything is the same except that the direction of Berry curvature is reversed, which we drop for clarity. The magnetic fluxes are always straight lines which emanate from the origin $O$ (the tip of the cone) in $(m_1,m_2,m_3)$ space as parameterized in Eq.~(\ref{decomposition}). $\theta$ introduced in Eq.~(\ref{density}) is the angle spanned by $m_3$ axis and direction of Berry curvature under study. There is no magnetic flux outside of the cone; in the cone the magnetic field becomes stronger as approaching the cone's surface and tends to infinity on the surface. Because the flux density assumes the axial symmetry about the $m_3$ axis, the two dimensional plot is depicted for clarity. }
\end{figure}

\section{Physical examples}
\label{examples}
In previous sections, we have developed and studied the Lorentz quantum mechanics for the simplest Lorentz spinor. Such a Lorentz spinor can arise in physical systems containing bosonic Bogoliubov quasiparticles, for example, in Bose-Einstein condensates(BECs)~\cite{njp}.
Specifically,  we illustrate our study of Lorentz quantum mechanics by investigating a 1D fermion gas at low temperatures, phonon excitations on top of a vortex in the BEC,  and spin wave excitations in a 1D antiferromagnetic system.

\subsection{One dimensional Fermi gas}

As the first illustrative example, we investigate the fermion excitations in a one dimensional fermion gas at low temperatures. Since excitations dominantly occur for fermions near the Fermi surface (note at 1D, the Fermi surface shrinks to the left (L) and right (R) Fermi points), the corresponding Hamiltonian can then be written as~\cite{fergas}
\begin{equation}\label{eq:HF}
H_{\textrm{F}}=\sum_{s=R,L}\sum_{q}(a_{sq}^\dag v_F q a_{sq}\kappa_s +\frac{1}{2N} g_4\rho_{sq}\rho_{s-q}+g_2\rho_{sq}\rho_{\bar{s}-q}).
\end{equation}
Here, the operator $a_{sq}^\dag$ ($a_{sq}$) creates (annihilates) an excited fermion near the Fermi point ($s=\textrm{R}, \textrm{L}$) with momentum $q$ (measured with respect to the ground state value). In addition, $\kappa_s= 1,-1$ for $s=\textrm{R}/\textrm{L}$, $\bar{s}=L/R$, $v_F$ labels the fermi velocity, and $\rho_{sq}=\sum_k a_{sk+q}^\dag a_{sk}$ is the density operator in the momentum space representation. In writing down Eq. (\ref{eq:HF}), we have taken into account the interactions between two fermions. Specifically, $g_2$ denotes the strength of interaction between two fermions near opposite Fermi points (i.e. $q\simeq2k_F$), while $g_4$ for those close to the same Fermi point (i.e. $q\simeq0$).

Let $|0\rangle$ denote the state of perfect Fermi sphere (a Fermi line in one dimensional case). A generic state describing density fluctuations near the Fermi points can then be written in terms of a pseduo-spinor as
\begin{equation}\label{eq:spinor1}
\left(\begin{array}{c}a \\b\end{array} \right)\equiv\frac{1}{\rho}\left(a \sqrt{\frac{2\pi}{lq}}\rho_{Lq}+b\sqrt{\frac{2\pi}{lq}} \rho_{Rq}\right)|0\rangle,
\end{equation}
where $l$ is the size of the system. As discussed in Ref. \cite{fergas}, the density operators $\rho_{sq}$ can be effectively treated as bosonic operators within the approximation
\begin{equation}\label{eq:density}
[\rho_{sq},\rho_{s'q'}]\simeq\langle 0 [\rho_{sq},\rho_{s'q'}] |0\rangle.
\end{equation}
By assuming Eq. (\ref{eq:density}), it is found that Eq. (\ref{eq:spinor1}) represents a Lorentz spinor whose dynamics is governed by the BdG equation below
\begin{equation}\label{eq:ex1}
{\text i}\hbar\frac{d}{dt}\left(\begin{array}{c}a \\b\end{array} \right)=\sigma_{1,1} q \left(\begin{array}{cc}v_F+\frac{g_4}{2\pi}&\frac{g_2}{2\pi} \\ \frac{g_2}{2\pi}&v_F+\frac{g_4}{2\pi}\end{array} \right) \left(\begin{array}{c}a \\b\end{array} \right).
\end{equation}

The generator $\sigma_{1,1}H$ of the dynamics in Eq. (\ref{eq:ex1}), when written in form of Eq. (\ref{decomposition}), corresponds to $m_1=g_2q/(2\pi)$, $m_2=0$ and $m_3=v_F q+g_4/(2\pi)q$. Thus, when $v_F+\frac{g_4}{2\pi}\geq\frac{g_2}{2\pi}$ [see Eq. (\ref{first-condition})], the $\sigma_{1,1}H$ exhibits real eigenvalues, and has a space-like and a time-like eigenvectors. Due to $m_2=0$, as illustrated in Fig.~3, there is no magnetic flux penetrating a loop in the plane defined by $(m_1,m_3)$. As a result, the Berry phase picked up by the eigenstate, say $|1(\mathbf{R})\rangle$, is always zero when $\mathbf{R}$ varies along a loop in the parameter space of $(m_1,m_3)$. According to our theory, it is impossible to implement a geometric force (vector potential or artificial magnetic field) to any fermions in the one dimensional Fermi gas. We must search for other intriguing systems to implement an artificial magnetic field. Below is an example.

\subsection{Phonon excitations on top of a Bose-Einstein condensate vortex}

The above example shows that the existence of a non-zero Berry phase requires $\sigma_{1,1}H$ - when written in form of (\ref{decomposition}) - to contain a complex part, i.e., $m_2\neq0$. Below, we demonstrate that this can be realized in the dynamics of phonons excited on top of a vortex in a BEC.

Following Ref.  \cite{ZhangNiu}, we assume the phonon wave packet has a narrow width smaller than all the relevant length scales associated with slowly varying potentials (e.g., trapping potential). The corresponding effective BdG equation can be derived as,
\begin{eqnarray}
{\text i}\hbar\frac{d}{dt}\left(\begin{array}{c}a \\b\end{array} \right)=\sigma_{1,1} \left(\begin{array}{cc}H_+&H_2 e^{2{\text i}\alpha(\mathbf{r})} \\ H_2 e^{-2{\text i}\alpha(\mathbf{r})}&H_-\end{array} \right) \left(\begin{array}{c}a \\b\end{array} \right), \label{eq:Phonon}
\end{eqnarray}
where $H_2=gn(\mathbf{r}_c)$ and
\begin{equation}
H_{\pm}=\frac{\mathbf{q}^2}{2}+2gn(\mathbf{r}_c)+V(\mathbf{r}_c)-\mu\mp \mathbf{\Omega}\cdot(\mathbf{r}_c\times\mathbf{q})\,.
\end{equation}
Here, $\mathbf{r}_c$ labels the coordinate of the vortex center, $g$ is the interatomic coupling constant, $V(\mathbf{r}_c)$ is the trapping potential of BEC, and $\mathbf{\Omega}$ is the rotating frequency of the whole system. Furthermore, $n(\mathbf{r_c})$ and $\alpha(\mathbf{r_c})$ denote the particle density and phase of the wavefunction around the vortex center, respectively, with $\mathbf{q}$ labeling the wave vector of phonons.

For every value of $(\mathbf{q},\mathbf{r}_c)$, the $\sigma_{1,1}H$ read off from Eq. (\ref{eq:Phonon}) can be cast into the form (\ref{decomposition}) with
\begin{eqnarray} \nonumber
m_1&=&gn(\mathbf{r}_c)\cos[2\alpha(\mathbf{r}_c)], \\ \nonumber
m_2&=&gn(\mathbf{r}_c)\sin[2\alpha(\mathbf{r}_c)], \\
m_3&=&\mathbf{q}^2/2+2gn(\mathbf{r}_c)+V(\mathbf{r}_c)-\mu.
\end{eqnarray}
In this case, the space-like eigenstate of $\sigma_{1,1}H$ reads
\begin{equation}\label{eq:eig1}
|1\rangle=\frac{1}{2}\left(\begin{array}{c}\zeta+\zeta^{-1} \\ (\zeta-\zeta^{-1})e^{-2{\text i}\alpha(\mathbf{r}_c)}\end{array} \right),
\end{equation}
with $\zeta=\left(\frac{H_1-m_3}{H_1+m_3}\right)^{1/4}$. The eigenstate (\ref{eq:eig1}) features a complex angle. As a result, when $\mathbf{r}_c$ varies in the real space, the eigenstate $|1\rangle$ will pick up a non-zero Berry phase: calculating the Berry connection
\begin{eqnarray} \nonumber
\mathbf{A}_1={\text i} \langle 1|\sigma_{1,1} \frac{\partial}{\partial \mathbf{r}_c}|1\rangle,
\end{eqnarray}
we derive the Berry phase as
\begin{eqnarray}
\beta_1=\oint d\mathbf{r}_c\cdot\mathbf{A}_1=-\oint(M-1)d\alpha(\mathbf{r}_c),
\end{eqnarray}
with $M$ the total atomic mass contained in the quasiparticle wave packet. The Berry connection $\mathbf{A}_1$ will then give rise to an effective vector potential (magnetic field) acting on the spatial motion of the vortex. In a previous study of the system \cite{ZhangNiu}, the vector potential has been worked out for a regime of the parameter space but the global feature of the distribution of the Berry-like curvature (magnetic field) is still left unknown. In our calculation, the distribution of magnetic field for the two-mode BdG equation is globally depicted in Fig.~3.

\subsection{Spin-wave excitations in antiferromagnet}

Here we demonstrate the Lorentz spin-orbital coupling (SOC) for the spin wave excitations in a 1D antiferromagnet. Concretely, we consider two sublattices, labeled by A and B, which encode the positive and negative magnetic moments near zero temperature. The corresponding Hamiltonian in the standard Heisenberg's description reads
\begin{eqnarray} \nonumber
&&H_s =J\sum_{i,\delta}[S_{ai}^zS_{b,i+\delta}^z+\frac{1}{2}(S_{ai}^+S_{b,i+\delta}^-+S_{ai}^-S_{b,i+\delta}^+)] \\
&&+J\sum_{j,\delta}[S_{bj}^zS_{a,j+\delta}^z+\frac{1}{2}(S_{bj}^+S_{a,j+\delta}^-+S_{bj}^-S_{a,j+\delta}^+)].\label{eq:Hspin}
\end{eqnarray}
where $\delta=\pm 1$ stands for the nearest neighboring sites, $J>0$ is the antiferromagnetic exchange integral, $S^z_{ai}$ ($S^z_{bj}$) are the spin operator (z component) on the sublattice A(B), and $S^\pm$ is the standard spin flip operators. Without loss of generality, we suppose the spins in the sublattice A (B) are along the positive (negative) $z$ direction in the limit of low temperatures.

Hamiltonian (\ref{eq:Hspin}) can be recast into a more transparent form using the Holstein-Primakoff transformation \cite{Holstein}. Briefly, introducing $a_i^\dag=S_{ai}^-$, and $b_i^\dag=S_{bi}^+$, together with the Fourier transformation into the momentum space,
\begin{eqnarray}
&&a_i=N^{-\frac{1}{2}}\sum_k e^{{\text i}kR_i}a_k, \quad a_i^\dag=N^{-\frac{1}{2}}\sum_k e^{-{\text i}kR_i}a_k^\dag, \\
&&b_j=N^{-\frac{1}{2}}\sum_k e^{-{\text i}kR_j}b_k, \quad b_j^\dag=N^{-\frac{1}{2}}\sum_k e^{{\text i}kR_j}b_k^\dag,
\end{eqnarray}
we rewrite Eq. (\ref{eq:Hspin}) as (dropping a constant)
\begin{eqnarray}  \label{HamiltonianK}
\tilde{H}_s&=&2ZSJ\sum_k(a_k^\dag a_k+b_k^\dag b_k+\gamma_k a_k^\dag b_k^\dag+\gamma_k b_k a_k) \nonumber\\
&=& 2ZSJ\sum_k \left(\begin{array}{cc}a_k^\dag&b_{k}\end{array} \right) \left(\begin{array}{cc}1&\gamma_k\\  \gamma_k&1\end{array} \right) \left(\begin{array}{c}a_k\\b_{k}^\dag\end{array} \right).
\end{eqnarray}
Here, $Z=2$ is the coordination number for the 1D system; $\gamma_k=\frac{1}{Z}\sum_\delta e^{{\text i}\mathbf{k}\cdot\mathbf{\delta}}=\cos(k)$ is the structure factor of the 1D lattice (here the lattice constant is taken as $a_l=1$, and the momentum is measured in the unit of $\hbar/a_l$). Let the ground state of Hamiltonian (\ref{HamiltonianK}) be denoted as $|0\rangle$, (which involves a superposition of enormous number of Fock states in the particle number representation $a_k^\dag a_k$, $b_k^\dag b_k$. )

The above Holstein-Primakoff transformation allows a vivid description of the spin wave excitations of the system [see Eq. (\ref{eq:Hspin})] in terms of ``particles" and ``holes" created in the ground state. In the simplest case, we consider the dynamics of an arbitrary (1,1)-spinor state given by
 \begin{equation} \label{representation}
\left(\begin{array}{c}a \\b\end{array} \right)\equiv\frac{1}{\rho}(a a_k^\dag+b b_{k})|0\rangle,
\end{equation}
with $\rho$ the normalization constant, corresponding to creations of a pair of particle and hole. The time evolution of Eq. (\ref{representation}) can be derived as
\begin{equation}\label{eq:spinor}
{\text i}\hbar\frac{d}{dt}\left(\begin{array}{c}a \\b\end{array} \right)=\sigma_{1,1} \left(\begin{array}{cc}1&\gamma_k \\ \gamma_k&1\end{array} \right) \left(\begin{array}{c}a
\\b
\end{array} \right),
\end{equation}
which features a $k$-dependent generator. The corresponding eigenspinors  $(u,v)^T$ and $(v^*,u^*)^T$ are found to be real and take the form
\begin{eqnarray}  \label{eigensolution}
u(k)&=&\sqrt{\frac{1}{2}\left(\frac{1}{|\sin(k)|}+1\right)}\,, \\
v(k)&=&{\text {sgn}}(\cos(k))\sqrt{\frac{1}{2}\left(\frac{1}{|\sin(k)|}-1\right)}\,,
\end{eqnarray}
which manifestly exhibit SOC effect, with the orbital state $k$ coupled to a Lorentz spinor. Since the SOC effect for the conventional unitary quantum mechanics has been studied extensively in both single-body systems~\cite{ZB,Vaishnav,Ruseckas,Juzeliunas}, where Zitterbewegung oscillation occurs~\cite{ZB,Vaishnav} and BEC systems~\cite{Zhai}, where single plane wave phase and standing wave phase were found, along this direction we may expect and explore the ample physical consequences of the Lorentz SOC.

\section{Conclusion}

To summarize, we have studied the dynamics of bosonic quasiparticles based on BdG equation
for the $(1,1)$-spinor. We show that the dynamical behavior of these bosonic quasiparticles is described by Lorentz quantum mechanics, where both time evolution of a quantum state and the representation transformation represent Lorentz transformations in the complex Minkowski space. The basic framework of the Lorentz quantum mechanics for the Lorentz spinor is presented, including construction of basic operators that are analogue of Pauli matrices. Based on it, we have demonstrated the Lorentz counterpart of the Berry phase, Berry connection, and Berry curvatures, etc. Since such Lorentz spinors can be generically found in physical systems hosting bosonic Bogoliubov quasi-particles, we expect that our study allows new insights into the dynamical properties of quasiparticles in diverse systems. In a broader context, the present work provides a new perspective toward the fundamental understanding of quantum evolution, as well as new scenarios for experimentally probing the coherent effect. While our study is primarily based on Bogoliubov equation for the $(1,1)$-spinor, we expect the essential features also appear in dynamics described by the Bogoliubov equation of multi-mode, the study of which is of future interest.

\end{document}